\documentclass[aps,prl,twocolumn,nofootinbib]{revtex4}
\usepackage{epsf,epsfig,amsmath,amssymb,dsfont}
\usepackage{aas_macros}
\newsavebox{\ns}
\newsavebox{\dbrane}

\def\be{\begin{equation}}
\def\ee{\end{equation}}
\def\ben{\begin{equation*}}
\def\een{\end{equation*}}
\def\bea{\begin{eqnarray}}
\def\eea{\end{eqnarray}}

\def\Dslash{\,\,{\raise.15ex\hbox{/}\mkern-12mu D}}
\def\Dbarslash{\,\,{\raise.15ex\hbox{/}\mkern-12mu {\bar D}}}
\def\delslash{\,\,{\raise.15ex\hbox{/}\mkern-9mu \partial}}
\def\delbarslash{\,\,{\raise.15ex\hbox{/}\mkern-9mu {\bar\partial}}}
\def\pslash{\,\,{\raise.15ex\hbox{/}\mkern-9mu p}}
\def\calDslash{\,\,{\raise.15ex\hbox{/}\mkern-12mu {\cal D}}}

\newcommand{\bbC}{\mathbb{C}}
\newcommand{\bbP}{\mathbb{P}}

\newcommand{\bbM}{\mathbb{M}}

\begin{document}

\title{Polarization Diffusion from Spacetime Uncertainty}

\author{Carlo R. Contaldi, Fay Dowker and Lydia Philpott}
\affiliation{Blackett Laboratory,
  Imperial College, London SW7 2AZ, U.K.}

\begin{abstract}
A model of Lorentz invariant random fluctuations in photon 
polarization is presented. The 
effects are frequency dependent and affect the polarization of photons
as they propagate through space. We test for this effect by
confronting the model with the latest measurements of polarization of
Cosmic Microwave Background (CMB) photons. 
\end{abstract} 

\maketitle


All approaches to the problem of quantum gravity 
predict that the spacetime itself will suffer from quantum 
uncertainty
at Planckian scales. This basic idea is
the inspiration behind many attempts to
formulate phenomenological models of quantum gravitational
effects with 
potentially observable consequences. 
To date, much of the effort spent on modeling the effect of 
spacetime fluctuations has produced Lorentz
symmetry violating models.
However, with constraints on violations of Lorentz symmetry becoming 
tighter all the time 
(see {\textit{e.g.}} \cite{Collaborations:2009zq})  
it is more important than ever to discover quantum gravity 
phenomenology that respects Lorentz symmetry. 

That such models can exist has been demonstrated in \cite{Dowker:2003hb} and 
\cite{Philpott:2008vd}. That work was motivated by the causal set 
approach to quantum gravity \cite{Bombelli:1987aa, Sorkin:2003bx, Henson:2006kf} 
but the scheme is quite general and does not depend on 
any details of the underlying theory except that it
should be Lorentz invariant. 
The basic idea is that 
certain dynamical quantities such as particle trajectories 
are subject to minute, quasi-local, random fluctuations due to 
the uncertainty in spacetime structure at the Planck scale. 
The model-building strategy is 
straightforward: identify a space of states for the 
system, work out how Lorentz transformations act
and hence deduce the most general 
Lorentz invariant diffusion process on that space.

In the case of a massive point particle, the outcome of this 
strategy is an Ornstein-Uhlenbeck-type process in which the 
momentum of the particle undergoes Brownian motion on the 
mass shell in proper time  \cite{Dowker:2003hb}.  
For massless particles, the Lorentz symmetry
restricts the process to be a one dimensional diffusion in 
energy -- the particles always travel on the light cone -- but a
second independent parameter enters which governs a
drift in energy \cite{Philpott:2008vd}. 
 
In this paper we will 
apply the 
strategy described above to polarization degrees of freedom 
as suggested in \cite{Sorkin:2007qi}. 
We model a photon classically as a point particle with 
a spacetime position $x^\mu$, null momentum $k^\mu = (k^0, \vec{k})$ and
a polarization state to be identified. 
A more realistic description would use wave packets 
and an even better model would take account of the quantal nature of 
photons. 
For now we assume that this classical state is 
a good approximate description of each of the free streaming 
photons produced by astrophysical and cosmological 
sources which reach our detectors. 

The state space for a classical photon is therefore
$\bbM^4 \times {\cal{H}}_0^3 \times \cal{B}$
where $\bbM^4$ is 4 dimensional Minkowski spacetime, ${\cal{H}}_0^3$
is the 3 dimensional ``cone'' of future-pointing null 4-vectors, and 
$\cal{B}$ is the space of polarization states which we 
will see is the Bloch sphere. 

The polarization state of a massless particle of momentum 
$k^\mu$ can be given by a complex 4-vector $a^\mu$ such that
$k^\mu a_\mu = 0$ and $a^{\mu*}a_\mu = 1$. 
The vector ${a'}^\mu = a^\mu + 
\lambda k^\mu$, for any complex number $\lambda$, will  
describe the same state. To eliminate this gauge freedom, 
we can consider the polarization state to be given by the complex
two form, $P = k \wedge a$, whose components, $P_{\mu\nu} = 
k_\mu a_\nu - a_\mu k_\nu$
satisfy the Lorentz invariant conditions
\bea
P^{\mu\nu}k_\nu&= 0\,,\\
P^{\mu\nu}P_{\mu\nu} &= 0\,,\\
P^{\mu\nu*}P_{\mu\nu} &= 0\,,\\
P^{\mu\nu*}P_{\mu\sigma} &= k^\mu k_\sigma\,.
\eea
If $k^\mu = s^\mu$ where $s^\mu := (1,0,0,1)$, $P$ has the 
following components
\be \label{PolMat.eq}
P_{\mu\nu} = \begin{pmatrix}
0 & -a_1 & -a_2 & 0 \\
a_1  & 0 & 0 & -a_1 \\
a_2 &0 & 0 & -a_2\\
0 & a_1 & a_2 & 0
\end{pmatrix}\,, 
\ee
where $a_1$ and $a_2$ are complex
numbers such that $|a_1|^2 + |a_2|^2 = 1$.
This corresponds to a polarization vector $a_\mu = (0,a_1,a_2,0)$.

The phase of the 2-d complex unit vector $(a_1, a_2)$ 
is not relevant for the polarization 
state of a single photon and so the 
polarization state space has two real dimensions: it is the 
Bloch sphere, ${\cal{B}} \cong \bbC {\bbP}^1 $. 
Let $\alpha$ and $\beta$ be, respectively,
the usual polar and azimuthal 
angles on $\cal{B}$, then they are related to the
components of $P_{\mu\nu}$ by
\bea \label{chipsi.eq}
a_1& = \frac{e^{i\gamma}}{\sqrt{2}} \left(\cos{\frac{\alpha}{2}}
+ e^{i\beta} \sin{\frac{\alpha}{2}}\right)\, ,\\
a_2 & = i\frac{e^{i\gamma}}{\sqrt{2}}\left( \cos{\frac{\alpha}{2}}
- e^{i\beta} \sin{\frac{\alpha}{2}}\right)\, ,
\eea
where $\gamma$ is an irrelevant phase. 
The 
north and south poles, 
$\alpha = 0,\pi$, are  
the circularly polarized states and the equator, $\alpha = \pi/2$,
consists of the linearly polarized states.

Now consider a general photon state $(k^\mu, P_{\mu\nu})$. 
For a general $k^\mu$, the polarization 2-form $P_{\mu\nu}$ must be 
transformed by a Lorentz transformation that takes $k^\mu$ to $s^\mu$ 
in order for it to be compared to the standard polarization basis and 
its coordinates on $\cal{B}$ determined. 
This can be done using a {\textit{standard Lorentz transformation}} 
defined for example in \cite{Weinberg:2005}.
If $P(k)$ is the polarization 2-form thus transformed, then 
it  will have components of the form (\ref{PolMat.eq}) and 
\be
P(k)_{\mu\nu} = s_\mu a_\nu - a_\mu s_\nu\,,
\ee
where $a_\mu = (0, P(k)_{10}, P(k)_{20}, 0)$.
The $(\alpha, \beta)$ coordinates of the polarization state 
on $\cal{B}$ are then obtained from 
(\ref{chipsi.eq}) with $a_1 = P(k)_{10}$ and $a_2 = P(k)_{20}$ . 

In this way, every photon state is specified by coordinates $(x^\mu, 
k^\mu, \alpha, \beta)$
on ${\mathbb{M}}^4 \times {\cal{H}}_0^3 \times \cal{B}$.  

Under a Lorentz transformation
the photon state $(k^\mu, P_{\mu\nu})$ transforms in
the usual way as a vector and 2-tensor and 
it can be shown that this translates into a
{\textit{polar rotation}} on $\cal{B}$, a rotation around
the north-south polar axis  
generated by $\frac{\partial}{\partial \beta}$. 
Details of these derivations will appear elsewhere. 

The Stokes parameters (see e.g. \cite{Kamionkowski:1996ks}) are a 
convenient way to parameterize the polarization of a beam of
electromagnetic radiation. 
A  monochromatic beam 
with 
Stokes parameters $(I, Q,U, V)$ can be modeled as a 
bunch of photons with the same momentum $k^\mu$ and
polarization states distributed over $\cal{B}$. $I$ is the 
intensity of the beam and since our process 
preserves particle number $I$ is fixed.  
If a beam consists of photons of momentum $k^\mu$ which are all in 
the same polarization state $(\alpha, \beta) \in \cal{B}$ then 
the Stokes parameters of this perfectly polarized beam are
$Q= I \sin\alpha\cos\beta$, 
$U= I \sin\alpha\sin\beta$ and 
$V= I \cos\alpha$. 

If the photons 
have a distribution of polarizations 
the Stokes parameters are weighted by the 
probability density $\rho(\alpha, \beta)$
on $\cal{B}$ {\textit{e.g.}} 
$Q= I \int_{\cal{B}}\,\sin\alpha\cos\beta\,\, \rho(\alpha, \phi)\, d \alpha d \beta$ and similarly for $U$ and $V$.
There are many 
distributions that will model a given set of Stokes 
parameters. For example, an
unpolarized beam, $Q= U =V = 0$,
could be modeled by a uniform distribution
of linearly polarized states, or a uniform distribution on the 
two circularly polarized states alone.
In general, the more spread out the distribution on $\cal{B}$, the
smaller the {\textit{polarization fraction}}, 
${\cal{P}}:= \sqrt{Q^2 + U^2 + V^2}/I$. 

Having identified the state space of the photon
as $ \bbM^4 \times {\cal{H}}_0^3 \times {\cal{B}}$
we can deduce
the most general Lorentz invariant 
diffusion process on this space. As described in \cite{Philpott:2008vd}
the trajectory in spacetime is simple: the photon moves along
null lines according to $\frac{d x^\mu}{d\lambda} = k^\mu$
where $\lambda$ is affine time. 
Moreover the process on the momentum
space ${\cal{H}}_0^3$ must not disturb the 
blackbody nature of the CMBR spectrum over the age of the 
universe. This means that we can neglect this effect for the 
purposes of this paper: we assume that the photon's frequency
is constant along its worldline. We are left with the 
task of deducing the Lorentz invariant
diffusion equation on $\cal{B}$.  

We refer to 
coordinates on $\cal{B}$ as $X^A = (\alpha, \beta)$.
Then, following \cite{Sorkin:1986}, the most general 
diffusion equation on $\cal{B}$ is
\begin{equation}
 \frac{\partial\rho}{\partial \lambda} =
 \partial_A\left(K^{AB}\,n\,\partial_B\left(\frac{\rho}{n}\right)-u^A\rho\right)\,,
 \label{e:diffequA}
\end{equation}
where $\lambda$ is affine time, 
$K^{AB}$ is a symmetric, positive semi-definite 2-tensor, 
$u^A$ is a vector and $n$ is a scalar density (``density of 
states'') on $\cal{B}$. These 
geometric quantities are 
the phenomenological parameters of the 
model and must be Lorentz invariant. 

There is an embarrassment of choice of parameters
because Lorentz transformations act as polar rotations
only. Any tensor, vector or scalar density
that does not depend on the azimuthal angle, $\beta$, 
is Lorentz invariant. Free parameters that are whole functions do not 
make for powerful phenomenology. If, however, we restrict attention 
to the {\textit{linear}} polarization states alone (corresponding
to setting Stokes parameter $V$ to zero), the 
model recovers its predictive power. 

The space of linearly polarized states is the unit circle, the 
equator of the Bloch sphere, and the Lorentz transformations 
act as rotations of the circle. The coordinate around the circle
is $\beta$ and there is, up to a constant factor, one
Lorentz invariant vector, $\partial/\partial \beta$. 
A Lorentz invariant density $n$ must be constant on 
the circle. We deduce a simple
diffusion-cum-drift on the circle for the 
distribution $\rho= \rho(\alpha, \beta)$:
\be
\frac{\partial \rho}{\partial \lambda} = 
  c \frac{\partial^2}{\partial \beta^2} \rho
- d \frac{\partial}{\partial \beta}\rho\,,
\ee
where $c>0$ and $d$ are constants.  

Transforming from affine time to ``cosmic time'', 
{\textit{i.e.}} time in the observatory frame, $t = h \nu \lambda$ 
where $h$ is Planck's constant and $\nu$ is the 
frequency of the photon, we have 
\be \label{diffusion.eq}
\frac{\partial \rho}{\partial t}
 =  \frac{c}{\nu} \frac{\partial^2}{\partial \beta^2} \rho
 - \frac{d}{\nu} \frac{\partial}{\partial \beta}\rho\, .
\ee
We absorbed the $h$ into the free parameters
governing the diffusion and drift. 
We see that the rates of  diffusion and drift in polarization angle 
are frequency dependent.
Note that the Lorentz symmetry we have assumed is only 
invariance under the proper orthochronous component of the 
full Lorentz Group. The drift term explicitly breaks parity 
invariance. 

Our model has assumed that spacetime is Minkowski spacetime. 
Following \cite{Philpott:2008vd} we can model the effect of
an expanding universe by setting the frequency 
to depend on time, $\nu = \nu(t)$, such that $a(t) \nu(t) = a_0 \nu_0$
where $a(t)$ is the scale factor of the universe, $a_0$ is the 
current value of $a$ and $\nu_0$ is the current (observed) 
value of the frequency of the photon. If we define  a new 
time coordinate $t'$ by $dt'/dt = a(t)/a_0$ then our 
diffusion equation keeps the same form as (\ref{diffusion.eq})
\be \label{diffusionprime.eq}
\frac{\partial \rho}{\partial t'}
 =  \frac{c}{\nu_0} \frac{\partial^2}{\partial \beta^2} \rho
 - \frac{d}{\nu_0} \frac{\partial}{\partial \beta}\rho\, .
\ee
For a matter dominated universe $a \sim t^{2/3}$ and 
a range of 
$t$ of $10^{60}$ Planck times becomes a range of $t'$ of 
$3/5 \times 10^{60}$ Planck times. We drop the subscript 
$0$ from $\nu$ and the prime from 
t in what follows.

Consider a beam of photons of frequency $\nu$ whose polarization is 
initially described by Stokes parameters $(U,Q)$.  
The drift will result, after a time $t$ in a beam whose
polarization angle, $\Phi$, has rotated by $\chi := td/\nu$:
\be 
\Phi = \tan^{-1}\left(\frac{U}{Q}\right) \rightarrow \Phi' = 
\Phi + \chi\,.
\ee
The diffusion will result in a decrease in the
magnitude of polarization by a factor $\exp(-\mu)$:
\be
{\cal{P}} := \sqrt{U^2 + Q^2} \rightarrow {\cal{P}}' = e^{-\mu} {\cal{P}}\,.
\ee
where $\mu$ can be calculated to be $4 t c/\nu$. 

Clearly, such an effect would be most pronounced in photons that have
propagated over long distances. Thus CMB photons, whose polarization
is correlated over incoming directions and have traveled over
cosmological distances without rescattering, offer the best chance to
constrain the parameters in the model. The polarization of the photons
is imprinted during the last stages of recombination as the photons
decouple from baryons and enter the free streaming regime. The
correlation in the polarization of photons arriving from different
directions is encoded in a set of angular power spectra $C_\ell^{\rm
  XY}$ where $\rm XY$ are the different spectral cross-correlation
in total intensity $T$ and grad-type and curl-like components $E$ and
$B$ respectively. The spectra are calculated by solving the full
Einstein-Boltzmann system describing the evolution of perturbed fluids
in a particular cosmological model \cite{Bond:1984}.  The rotation and
suppression of polarization along the trajectory will modify the
angular power spectra of the observed CMB photons $ C_\ell^{\rm XY}
\rightarrow \tilde C_\ell^{\rm XY}$. The mapping for each spectrum is
given by \cite{Lue:1998mq,Feng:2004mq}
\begin{eqnarray}\label{eq:cls}
\tilde C_\ell^{EE} &=& e^{-2\mu}\, C_\ell^{EE}\, \cos^2(2\chi)\,,\nonumber\\
\tilde C_\ell^{BB} &=& e^{-2\mu}\, C_\ell^{EE}\, \sin^2(2\chi)\,,\nonumber\\
\tilde C_\ell^{TE} &=& e^{-\mu}\, C_\ell^{TE}\, \cos(2\chi)\,,\\
\tilde C_\ell^{TB} &=& e^{-\mu}\, C_\ell^{TB}\, \sin(2\chi)\,,\nonumber\\
\tilde C_\ell^{EE} &=& \frac{1}{2}e^{-2\mu}\, C_\ell^{EE}\, \sin(4\chi)\,,\nonumber
\end{eqnarray}
where we have assumed no $BB$ contribution to the original spectra (no
primordial gravitational waves). The $TT$ spectrum is not modified as
it is not sensitive to the polarization of the photons.

A number of assumptions are implicit in the simple mapping given in
(\ref{eq:cls}). Firstly it assumes that the picture of recombination
(when the polarization is imprinted on the CMB) and reionisation (a
further source of polarization) is not altered in this model. It
assumes the background cosmological evolution is the same for a given
set of cosmological parameters. In fact, polarization on large scales
is also generated after the universe is reionized and this could
introduce a mild scale dependence of the diffusion-rotation effect.

To constrain this scenario with available CMB data we modify the {\tt
  CosmoMC}\footnote{http://cosmologist.info/cosmomc/} Monte Carlo
Markov Chain (MCMC) package to fit for standard $\Lambda CDM$ model
parameters together with polarization rotation $\chi$ and polarization
depth $\mu$. The standard parameters are: cold dark matter and
baryonic matter physical densities $\Omega_ch^2$ and $\Omega_bh^2$,
angular diameter distance measure $\theta$, optical depth to
reionisation $\tau$, and primordial scalar perturbation amplitude
$A_s$ and spectral index $n_s$. We assume a uniform prior of
sufficient range in each parameter and do not include any primordial
tensor contributions. However we {\sl do} fit to $TB$, $EB$, and $BB$
data since these are not expected to vanish any longer in the modified
model.

We fit to a combination of data which includes all polarization
sensitive experiments which have reported a detection of the $EE$
power. These are the DASI results \cite{Leitch:2005}, the final
CBIpol results \cite{Sievers:2005}, the Boomerang 2003 flight results
\cite{Montroy:2006}, the WMAP 5-year results \cite{Nolta:2009} and the
latest BICEP \cite{Chiang:2009} and QUaD \cite{quad} results. Of
these, the last two contain the highest signal-to-noise determination
of the polarization spectra on scales below a degree. Both BICEP and
QUaD have published $TB$ and $EB$ data which are crucial in
constraining polarization rotation effects
\cite{Chiang:2009,Xia:2009}. We include the published $TB$ and $EB$
band powers with band power window functions mimicking the published
$TE$ and $EE$ ones. The frequency dependence is accounted for by
scaling the effect for each experiment to a reference frequency of 150
GHz. 

\begin{figure}[t]
\begin{center}
\includegraphics[width=3.5in]{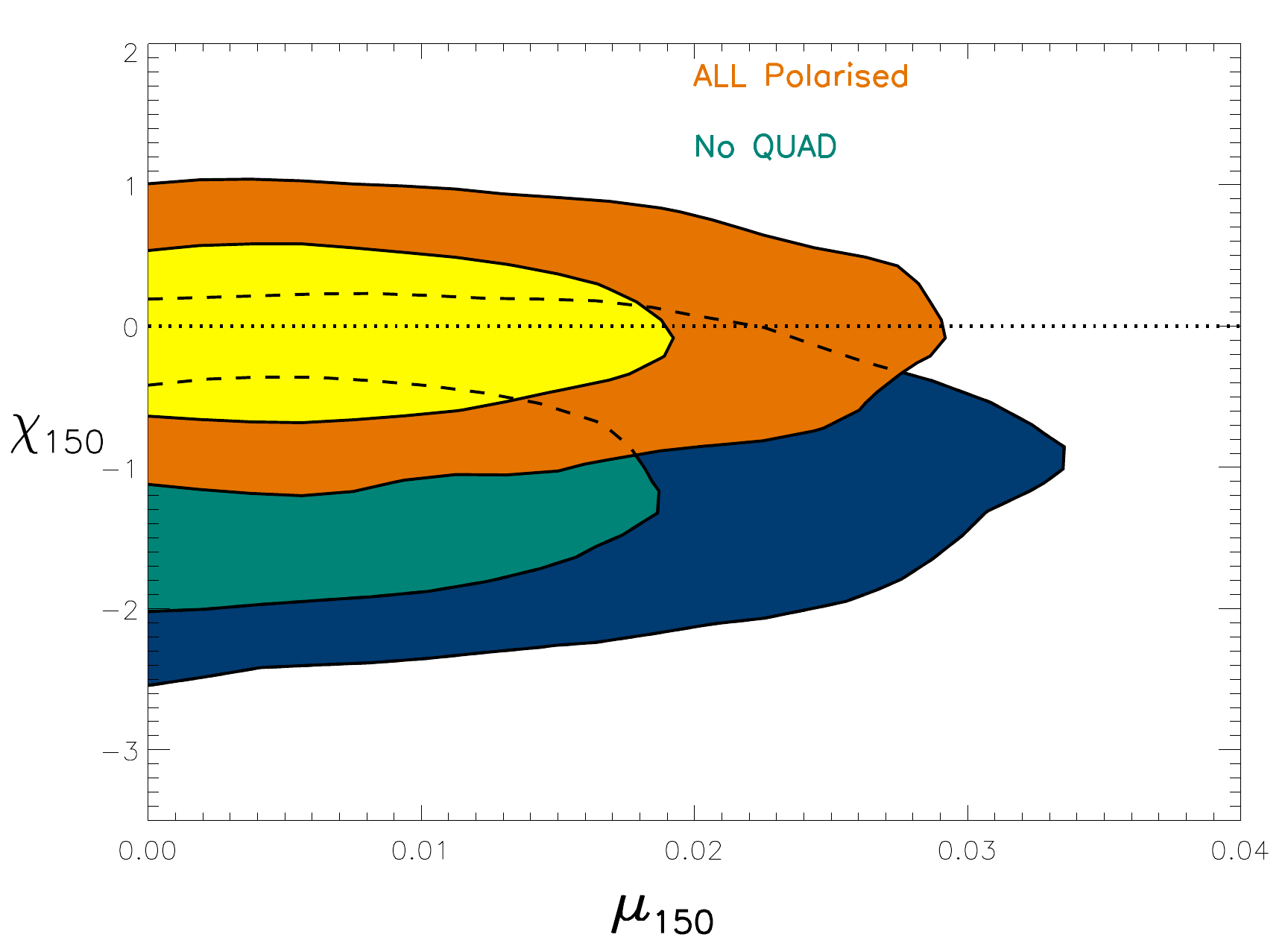}
\end{center}
\caption{The marginalized 2-d posterior density in the polarization
  rotation angle $\chi_{150}$ (in degrees) and polarization depth
  $\mu_{150}$ for the reference frequency of 150 GHz. The contours
  indicate the area bounding 68\% and 95\% of the density. We show the
  result for two data combinations; the first includes all
  polarization data and the second excludes the QUaD results. The
  results are consistent with no effect, however including the
  frequency dependence reduces the tension between the QuaD and
  `no QUaD' combinations highlighted in
  \cite{Xia:2009}.}\label{fig:res}
\end{figure}

The MCMC chains sample the posterior density in the 8-dimensional
parameter space. Once the sampling has converged we integrate the
densities over the standard parameters which gives the marginalized
posterior in $\chi_{150}$ and $\mu_{150}$. We show the result of the
marginalization in the $\chi_{150}$ {\sl} $\mu_{150}$ plane in
Fig.~\ref{fig:res}. We plot the 68\% and 95\% density contours for two
data combinations. The tightest constraints are obtained by the
combination of all polarized data and are consistent with no rotation
and vanishing polarization depth. We also include the results for the
case where the QUaD data is excluded. Although the `no QUaD' result,
driven mainly by the BICEP and WMAP measurements, prefers a non-zero
rotation angle, the indication is weaker than that reported in
\cite{Xia:2009}. However we {\sl do} recover their result when not
accounting for the frequency dependence. This may be an indication of
a frequency dependent effect in the data whereby the lower frequency
WMAP data tend favour less rotation given its effect is roughly
twice as large compared to the reference frequency of 150GHz. Future
observations at multiple frequencies will easily determine whether
this is the case.

\begin{table}
  \caption{Marginalized 1-d constraints for the polarization rotation
    angle $\chi$ and 95\% upper limits for the polarization depth
    $\mu$. }\label{tab:res}
  \begin{ruledtabular}
    \begin{tabular}{l|l|l}
      & ALL Pol & No QuaD\\
\hline
      Pol rotation: $\chi$ (degrees) & $-0.05^{+0.43}_{-0.43 }$& $-1.11^{+0.55}_{-0.55}$\\
      Pol depth: $\mu$ & $< 0.024 (95\%)$ &$< 0.026 (95\%)$
    \end{tabular}
  \end{ruledtabular}
\end{table}

Table~\ref{tab:res} shows the 1-d marginalized constraints on
$\chi_{150}$ and $\mu_{150}$ showing the marginal bias towards a
negative rotation driven mainly by the BICEP and WMAP data. The ``All
polarized'' result is consistent with no effect.


The model building strategy employed here is based on the assumption
that the fluctuations in dynamical variables due to spacetime
uncertainty are small enough that they result, in the hydrodynamic
approximation, in a continuous, Brownian motion through the state
space.  If spacetime uncertainty causes more violent, discontinuous
jumps in the variables, this will have to be modeled by a Boltzmann
equation rather than a diffusion equation.  For this reason and also
as data begins to constrain our phenomenology, the need for
microscopic models that can give us a handle on the parameters from
more fundamental physics becomes more acute.  For example, an
improvement on point particle models would be to treat particles as
wave packets of a scalar field on a causal set using the discrete
D'Alembertian operators described in
\cite{Henson:2006kf,Sorkin:2007qi,Benincasa:2010ac}.  Assuming
continuity, however, the power of our model is its robustness: if
photons can be modeled as classical particles with polarization
degrees of freedom, then \emph{any} Lorentz invariant effect of
underlying spacetime uncertainty -- whether due to discreteness,
fluctuations, fuzziness, foaminess or whatever -- will be of this
form.

\vspace{-0.3cm}
\begin{acknowledgments}
We thank Rafael Sorkin for helpful discussions. FD's research was 
supported by EC grant MRTN-CT-2004-005616, 
and Royal Society Grant IJP 2006/R2. 
LP acknowledges the support of a TEC doctoral scholarship.
FD is grateful to the
Perimeter Institute for Theoretical Physics, Waterloo, Canada, for hospitality
during work on this paper.
\end{acknowledgments}


\vspace{-0.7cm}

\bibliographystyle{apsrev}
\bibliography{}

\end{document}